\documentclass[11pt,twoside
]{article}

\usepackage{baaa2011}
\usepackage{graphicx}
\usepackage{subfigure}
\usepackage{psfrag}
\usepackage{amssymb}
\usepackage[spanish,activeacute,english]{babel}
\usepackage[latin1]{inputenc}
\usepackage[T1]{fontenc} 
\usepackage{ae,aecompl} 
\usepackage{latexsym}
\usepackage{verbatim}
\usepackage{amsmath}
\usepackage{stmaryrd}
\usepackage{amsfonts}
\usepackage{amssymb}
\usepackage{wasysym}
\usepackage[colorlinks=true,dvips]{hyperref}

\begin{document}
\myselectenglish
\vskip 1.0cm
\markboth{R. Barb\'a et al.}%
{NIP of Stars}

\pagestyle{myheadings}
\vspace*{0.5cm}
\noindent PRESENTACIÓN MURAL
\vskip 0.3cm
\title{Southern near-infrared photometric monitoring of Galactic young
  star clusters ({\em NIP of Stars})}


\author{R. Barb\'a$^{1,2}$, N. Morrell$^{3}$, G. Gunthardt$^{2,4}$,
  S. Torres Robledo$^{2}$, M. Jaque$^{1,2}$, M. Soto$^{2}$,  G. Ferrero$^{5}$,
  J. Arias$^{2}$, A. Román-Lópes$^{2}$, R. Gamen$^{5}$, and J. Astudillo
  Hormazabal$^{2}$}

\affil{%
  (1) ICATE-CONICET, Av. España 1512 S, 5400 San Juan, Argentina\\
  (2) Departamento de Física, Universidad de La Serena, La Serena, Chile\\
  (3) Las Campanas Observatory, Colina El Pino, La Serena, Chile\\
  (4) Observatorio Astronómico de Córdoba, Universidad Nacional de C\'ordoba,
  Laprida 854, 5000 C\'ordoba, Argentina\\  
  (5) Facultad de Ciencias Astronómicas y Geofísicas, Universidad Nacional de
  La Plata, Paseo del Bosque S/N, 1900 La Plata, Argentina\\
}

\begin{abstract} 
We have performed a near-infrared photometric monitoring of 39 galactic young
star clusters and star-forming regions, known as {\em NIP of Stars}, 
between the years 2009--2011, using the Swope telescope at Las Campanas
Observatory (Chile) and the RetroCam camera.  
The primary objective of the campaign is to perform a census of photometric
variability of such clusters and to discover massive eclipsing binary
stars. In this work, we describe the general idea, the implementation of
the survey, and the first preliminary results of some of the observed
clusters. 
This monitoring program is complementary to the {\em Vista Variables
  in the Vía Láctea} ({\em VVV}), as the brightest sources observed in 
{\em NIP of Stars} are saturated in {\em VVV}. 
\end{abstract}

\begin{resumen}
Hemos realizado un monitoreo fotométrico infrarrojo de 39 cúmulos jóvenes y
regiones de formación estelar galácticas, conocido como {\em NIP of Stars},
entre 2009--2011,
utilizando el telescopio Swope del Observatorio las Campanas (Chile) y la
cámara RetroCam.  
El objetivo primario de la campaña es hacer un censo de variabilidad
fotométrica de tales cúmulos, y descubrir binarias masivas eclipsantes. 
En este trabajo presentamos la idea general del proyecto la
implementación de un datoducto semiautomático, como así también
resultados preliminares en uno de los cúmulos observados. 
Este programa de monitoreo es complementario al relevamiento {\em Vista
  Variables in the Vía Láctea} ({\em VVV}), dado que las fuentes más brillantes
 observadas en {\em NIP of Stars} saturan en {\em VVV}.
\end{resumen}

\section{Motivation and sample selection}

\noindent The mass, mass-loss and rotation are among the most important
parameters that govern the stellar evolution. In the context of stellar
masses, massive O and WR-type binaries are key objects because they enable us
to determine minimum masses from the solution of their radial-velocity (RV)
curves, and in the case to constrain the orbital inclination (for example
through eclipsing binaries), we can get absolute masses.
Knowing the multiplicity of massive stars is important because this factor has
a deep impact on stellar evolution, the initial mass function, and on the
energy balance of its environment and helping with clues about their origins
(Zinnecker \& Yorke, 2007). 

The ample majority of studies about massive eclipsing binaries comes from
observations in  the optical range, therefore the sample is limited to very
few objects with relatively low reddening. 
Surprising to learn that of the 370 O-type stars which are counted in the 
{\em The Galactic O Star Catalog} ({\em GOS}, Maíz Apellániz et al. 2004), 
there are only 38 eclipsing or ellipsoidal variables. 
From this group, no more than fifteen systems have reliable light- and RV-
curves. 
The panorama resulting in the infrared is much worse:
there are only five massive eclipsing systems with published data.
Conclusion: all our knowledge about the absolute masses of massive stars 
of O- and WN-type is derived from only few tens of objects. A situation which
poses challenges.

The primary objective of this project is to conduct a census of photometric
variability in a set of young galactic open clusters and star forming regions
affected by large extinction ($A_{\rm V} = 6 -30$). From those variable stars,
we are specially interested in the massive eclipsing binaries, which can be 
observed spectroscopically to determine absolute stellar parameters.
We have selected thirty-nine galactic young clusters and star-forming regions
following these criteria: 
a) clusters must be more or less resolved at scale
of one arcsecond, with uncrowded background to get reliable photometry. Thus,
clusters like Arches are discarded; 
b) some of its massive members must have spectral classification;
c) previous studies must have indication of the presence of at least
  five stars with spectral type earlier than B0;
d) such stars must be in the $H$-magnitude range $8 <H< 12$.

This NIR photometric monitoring program is very complementary to the  
{\em Vista Variable in the Vía Láctea} ({\em VVV}, Minniti et al. 2010)
survey, as the brightest sources of {\em NIP of Stars} are saturated in {\em
  VVV} images.

\section{Observing campaigns and pipeline}

The observations were carried out using the RetroCam camera attached to the
Swope 1-meter telescope at Las Campanas Observatory (Chile) during three
seasons in 2009 to 2011.   
Thirty-eight observing nights presented photometric conditions, from a total
of seventy-three nights. Ten nights were completely lost due to bad weather. 
The RetroCam camera (Hamuy et al. 2006) consists of an one-megapixel
Rockwell Hawaii-1 HgCdTe array, with a spatial scale of $0\farcs54$ per
pixel, which provides a $9'\times9'$ field-of-view (FOV). 
This spatial resolution is about four times better than Two-Micron All Sky
Survey images ({\em 2MASS}, Cutri et al. 2003). The monitoring campaign was
performed preferentially in the $H_{\rm C}$ filter, and occasionally in $J_{\rm
  S}$ and $Y_{\rm C}$, as this camera does not have $K$-band filter. 

For the reduction of hundreds of thousands of observations, we have
implemented a semi-automated pipeline, which is based in part on the
procedures used in the {\em Carnegie Supernova Project} ({\em CSP}, Hamuy et
al. 2006).  
The requirements of our project are much more severe than {\em CSP} in terms of
background subtraction in areas with high-density of stars and very bright
nebulosities. 
The pipeline is based on a series of {\em IRAF} scripts, shell
scripts in {\em gawk}, {\em FORTRAN} code, and makes use of {\em SExtractor}
code (Bertin \& Arnouts, 1996). 
Furthermore, it is structured in {\em Python} programming language. 
In a second stage, we plan to obtain astrometric solutions using {\em Swarp}
and {\em Scamp} codes (Bertin, 2006), and the photometric zero-points using 
{\em 2MASS} and {\em VVV} surveys. 
Reduced images and metadata are being stored in a database managed by {\em
 MySQL}. Figure 1 shows an example of $H_{\rm C}$ images.

\begin{figure}[!t]
\centering
  \includegraphics[width=.25\textwidth]{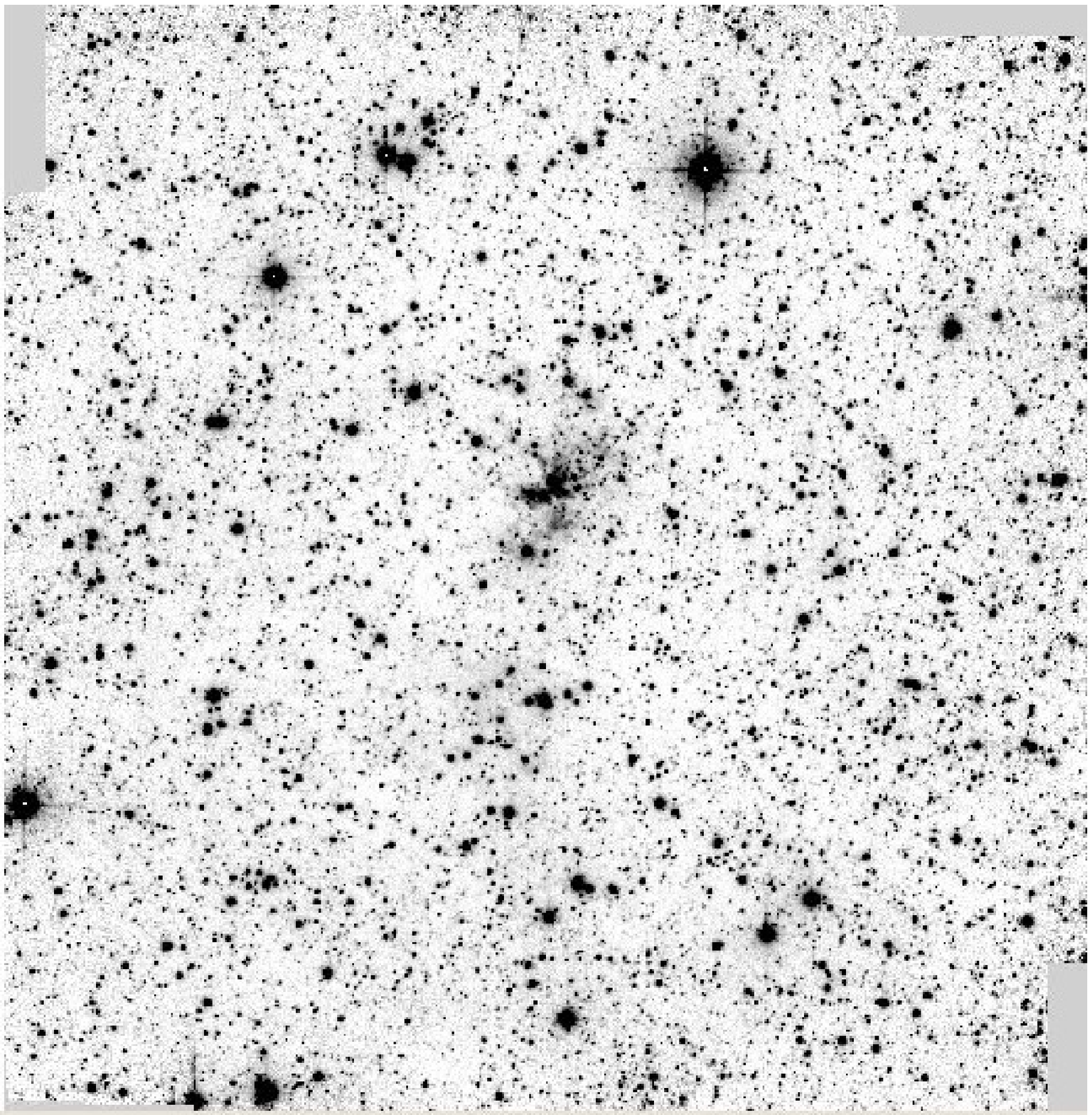}
  \includegraphics[width=.25\textwidth]{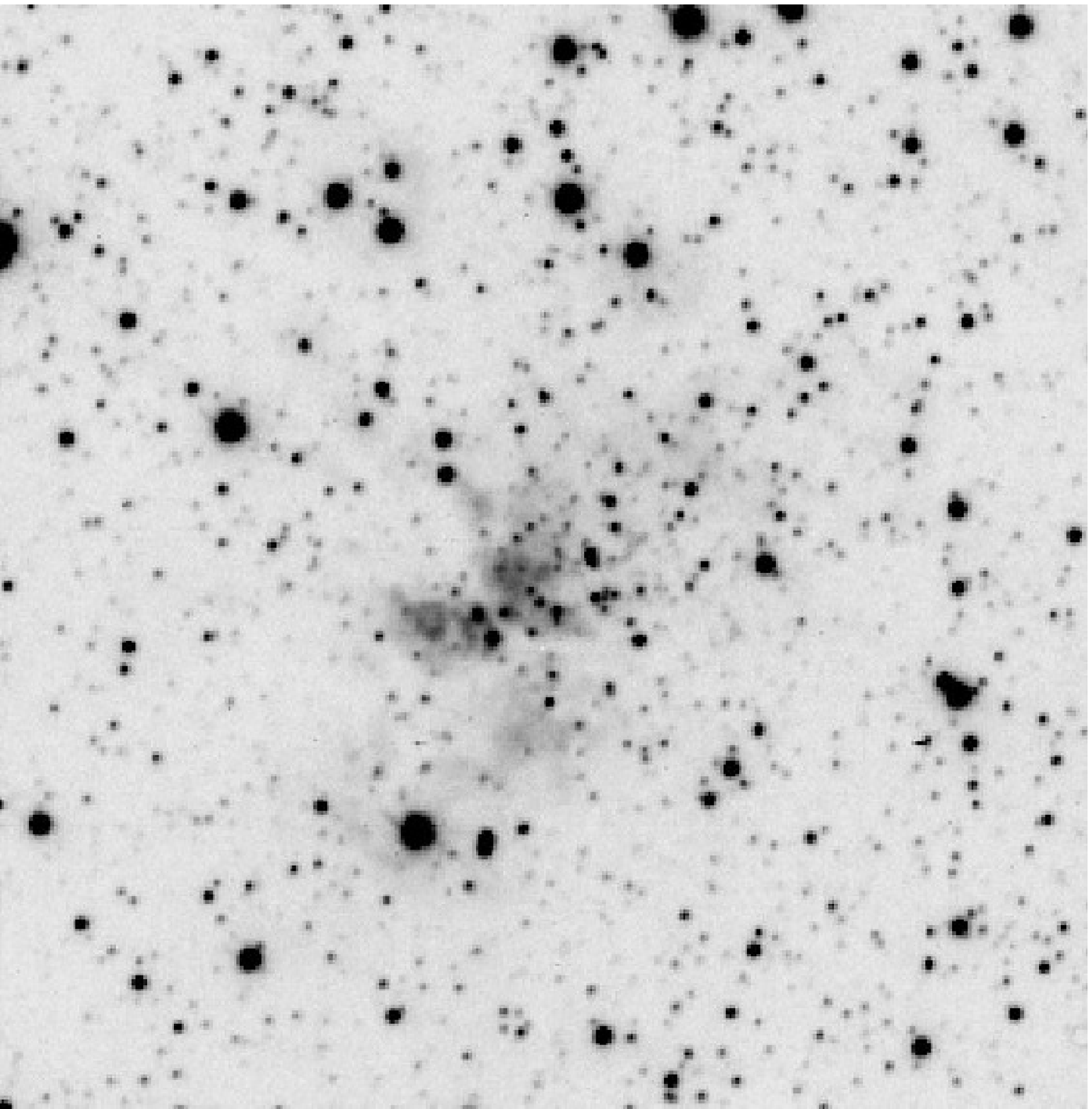}
\caption{Left. RetroCam $H_{\rm C}$ image of the star-forming region
  IRAS 16177--5018 obtained in July 2009, the FOV is about $10'$. 
  This is an example of an infrared cluster with relatively uncrowded
  background. 
  Right: the core of this cluster.}
  \label{fig:ab1}
\end{figure}

\section{First steps in the photometric analysis}

In a first stage, we are performing aperture photometry of a small set of
objects in order to check the photometric stability of observations. 
From this process we are getting excellent results, which guarantee relative
errors comparable to those obtained using well exposed CCD images in the 
optical range.  

Figure 2 shows some results of the differential photometry for five $H_{\rm
  C}$ mosaics of the young clusters Danks 1 and 2. This differential
photometry is relative to a $H_{\rm C}$ mosaic used as reference image. 
A mean-$\sigma$ of 0.02 mag is obtained for the instrumental magnitude
difference for twenty images of Danks 1 and 2, in the range $14-16$ mag 
(S/N$>200$). 
Thus, about 5\% of the sources show variability greater than $2\sigma$: 
these stars are potential variables. 
After this stage of photometric characterization of the sample, we plan to
design a pipeline to perform automated {\em point-spread-function} photometry
and we will start to test the {\em image-subtraction algorithm},
(Alard \& Lupton, 1998).

As a pilot case, we have observed the massive eclipsing binary FO15
(O5.5\,V + O9\,V, $P=1.41$d) in the Carina Nebula, with the aim to evaluate
the quality of the differential photometry procedures.
Figure 3 shows the phased light-curve of FO15 in the $Y_{\rm C}$ band, which
can be compared with that published by Niemela et al. (2006) using optical
{\em All-Sky Automated Survey} ({\em ASAS}) (Pojma\'nski 2003) observations. 
In spite of the fact that aperture photometry of FO15 was done without the
appropriate photometric calibrations, it is clear the superior quality of the
NIR light-curve. 

{\bf Acknowledgments.} 
{\small We thank support from DIULS PR09101 and FONDECYT 3110188. 
We thank to Director and staff of LCO for the use of their facilities.}

\begin{figure}[!ht]
  \centering
  \includegraphics[width=.48\textwidth]{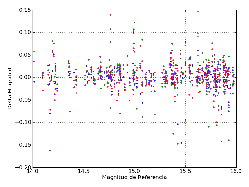}
  \includegraphics[width=.48\textwidth]{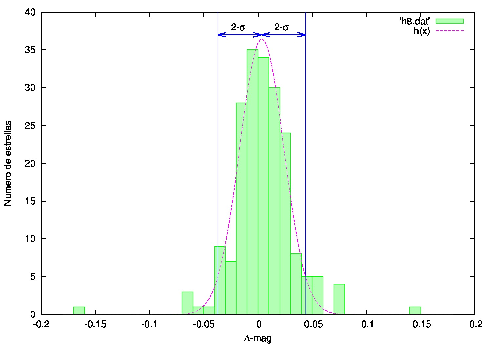}
  \caption{Left: differential instrumental magnitude for five $H_{\rm C}$
    mosaics of Danks 1 and 2 clusters respect to a reference mosaic. 
    The typical $\sigma$ for each star is about 0.02 mag. Sources with
    $\Delta H_{\rm C} > 2\sigma$ are potential variables.
    Right: distribution of $\sigma$ (green histogram) for the differential
    photometry of twenty $H_{\rm C}$ images of the same clusters. Variable
    star candidates are those beyond the $2\sigma$ interval. The fitted curve
    (in red) correspond to a Gaussian function with $\sigma=0.020$.}
  \label{fig:ab2}
\end{figure}

\begin{figure}[!ht]
  \centering
  \includegraphics[width=7.0cm]{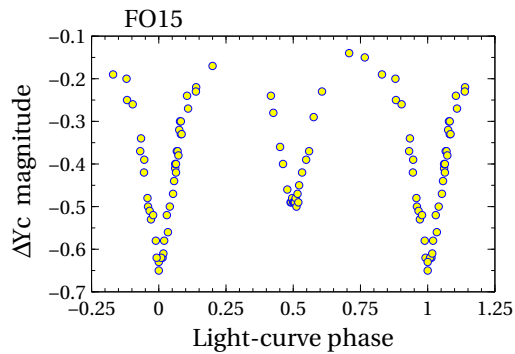}
  \caption{Near-infrared light-curve of the massive eclipsing binary FO15 in
    the Carina Nebula. The photometric data were phased using the
    ephemeris published by Niemela et al. (2006).}
  \label{fig:ab3}
\end{figure}
                                                                                
\begin{referencias}
\reference Alard, C. \& Lupton, R.H. 1998, \apj, 503, 325.
\reference Bertin, E. 2006, ASP Conf. Ser., Vol. 351, Eds.: C. Gabriel et al.,
112. 
\reference Bertin, E. \& Arnouts, S. 1996, Astron. Astrophys. Supp. Ser., 317,
393. 
\reference Cutri, R.M. et al. 2003, ``2MASS All-Sky Catalog of Point
Sources'', University of Massachusetts and Infrared Processing and Analysis
Center, Pasadena.
\reference Hamuy, M. et al. 2006, \pasp, 118, 2.
\reference Maíz Apellániz, J., Walborn, N.R., Galué, H., Wei, L.H. 2004,
\apjs, 151, 103.
\reference Minniti, D. et al. 2010, New Astronomy, 15, 433.
\reference Niemela, V.S. et al. 2006, \mnras, 367, 1450.
\reference Pojma\'nski, G. 2003, Acta Astron., 53, 341
\reference Zinnecker, H. \& Yorke, H.W 2007, \araa, 45, 481. 
\end{referencias}

\end{document}